\documentclass[openacc]{rstransa}%

\titlehead{Research}

\begin{document}

\title{Phase-drifting with emitting plasma temperature in the quasi-periodic pulsations of an X-class solar flare}

\author{%
Libo Fu$^{1}$, Valery M. Nakariakov$^{2,3}$, Ding Yuan$^{1}$, Suraj Sahu$^{1}$, Song Feng$^{4}$, Ehsan Tavabi$^{1}$}

\address{$^{1}$Key Laboratory of Solar Activity and Space Weather, School of Aerospace, Harbin Institute of Technology, Shenzhen, Guangdong 518055, China.\\

$^{2}$Centre for Fusion, Space and Astrophysics, Department of Physics, University of Warwick, Gibbet Hill Road, Coventry, CV4 7AL, United Kingdom.\\
$^{3}$Centro de Investigacion en Astronom\'ia, Universidad Bernardo O'Higgins, Avenida Viel 1497, Santiago, Chile.\\
$^{4}$ Faculty of Information Engineering and Automation, Kunming University of Science and Technology, Kunming 650500, China. 
}

\subject{solar system}


\corres{Ding Yuan\\
\email{yuanding@hit.edu.cn}}

\begin{abstract}
Recent multi-wavelength observations of solar flares have provided new constraints on the physical origin of quasi-periodic pulsations (QPPs). In an X-class flare, we detect a short-lived $\sim$5-minute QPP simultaneously in hard X-rays, extreme-ultraviolet (EUV), and soft X-ray emissions, exhibiting a clear phase-drifting behavior with emitting plasma temperature. Based on phase-resolved timing analysis, it is found that (i) the QPPs in all diagnostics share nearly identical oscillation periods, (ii) a systematic temperature-dependent phase drifting is present, with the phase delay relative to the hard X-ray emission increases systematically from the hottest to cooler EUV channels, and (iii) the QPP persists for only a few cycles during the impulsive phase. These properties imply that periodic magnetic reconnection, possibly triggered by the leakage of 5-minute oscillations from the lower atmosphere, modulates the non-thermal electrons responsible for the leading Hard X-ray QPPs. Subsequently, plasma heating and cooling processes manifest sequentially across passbands with different temperature responses, resulting in the observed temperature-dependent phase drifting. These results provide novel observational evidence supporting the use of multi-temperature, multi-wavelength phase relationships to constrain the temporal evolution of flare energy release and the origins of QPPs.
\end{abstract}

\begin{fmtext}

\end{fmtext}
\maketitle

\section{Introduction}

Quasi-periodic pulsations (QPPs) are a common and, perhaps, fundamental feature of solar and stellar flares, 
see \cite{2009SSRv..149..119N, 2016SoPh..291.3143V, 2019PPCF...61a4024N, 2020STP.....6a...3K, 2021SSRv..217...66Z} for comprehensive reviews. The QPPs appear as repetitive modulations of the solar and stellar flare emission, and have been reported across the entire electromagnetic spectrum, ranging from radio and microwave to white-light, UV/EUV, X-rays, and gamma rays, spanning both thermal and non-thermal emission components, e.g., \cite{2010SoPh..267..329K, 2012ApJ...749...28T, 2021ApJ...923L..33K}. 
The oscillation periods of QPPs span a vast range, from sub-second periods, typically in the radio and hard X-ray bands (see, e.g., \cite{2025MNRAS.542L..48L}), to tens of minutes in thermal emissions (e.g., \cite{2024A&A...690A..39L}). This suggests that QPPs are driven by different physical mechanisms or operate in diverse plasma environments. The qualifier \lq\lq quasi-periodic\rq\rq\ reflects the fact that these signals commonly exhibit pronounced temporal evolution of their amplitude, period, and phase, rather than behaving as strictly monochromatic sinusoidal oscillations with a constant period \cite{2019PPCF...61a4024N}. 
QPPs have been detected in flares over a broad range of energies, from microflares (e.g., \cite{2018ApJ...859..154N}) and small-scale energy release events up to the most powerful X-class flares (e.g., \cite{2020ApJ...893....7L, 2018ApJ...858L...3K, 2019ApJ...875...33H}). 
A recent work by Lim et al. (2025) \cite{2025A&A...698A..65L} further demonstrates that robust QPP signatures can be found even in small EUV brightenings that are considered to be candidate nanoflares, with characteristic periods comparable to those measured in \lq\lq classical\rq\rq\ solar flares of medium and large classes.

Multi-wavelength observations of solar flares provide a comprehensive view of energy release and particle acceleration across atmospheric layers \cite{2011SSRv..159...19F, 2017LRSP...14....2B}. Consequently, QPPs serve as a powerful diagnostic tool: their dependence on local plasma parameters allows one to distinguish between periodicities intrinsic to energy release (e.g., repetitive reconnection) and those arising from MHD wave modulation \cite{2009SSRv..149..119N,2021SSRv..217...66Z}. This seismological potential is particularly crucial for stellar flares, where the lack of spatial resolution leaves temporal signatures as the primary probe of flare physics (see, e.g., \cite{2016MNRAS.459.3659P}).

Estimates of QPP detection rates vary substantially depending on the instrumentation and methodology. While earlier studies or those using standard spectral techniques suggest a high prevalence of 70–90\% across flare classes \cite{2015SoPh..290.3625S,2018SoPh..293...61D,2025A&A...694A.251S}, rigorous Bayesian approaches that account for colored noise yield more conservative estimates of 20–46\% \cite{2020ApJ...895...50H,2016ApJ...833..284I}. However, these lower figures likely represent a lower limit, as strict detection criteria often overlook the non-stationary and anharmonic nature inherent to many QPP events \cite{2019PPCF...61a4024N,2023MNRAS.523.3689M}.

From a theoretical perspective, a large number of mechanisms have been proposed to explain the QPP modulation of flaring emissions; see \cite{2009SSRv..149..119N,2018SSRv..214...45M,2021SSRv..217...66Z} for detailed reviews. In general, QPP models can be broadly grouped into two (or three) partially overlapping categories: modulation of the magnetic reconnection rate or of the physical conditions in the emitting plasma by magnetohydrodynamic (MHD) oscillations, and  spontaneous repetitive reconnection. 

A promising way to distinguish between different theoretically possible mechanisms for QPPs is to search for differences in QPP parameters detected at different wavelengths. In the majority of cases, QPPs detected in different observational bands appear to be synchronous (see, e.g., \cite{2022Univ....8..358S, 2024A&A...684A.215C, 2024ApJ...970...77L} for recent results and references therein). Nevertheless, phase shifts between QPP patterns detected in different bands have also been observed.

The oscillation phase and the corresponding modulation amplitude of QPPs produced by sausage and kink oscillations of the emitting volume have been shown to depend on the radio wavelength (see, e.g., \cite{2012ApJ...748..140M, 2022MNRAS.516.2292K}). Dolla et al. \cite{2012ApJ...749L..16D} reported time lags of up to 9~s between QPPs observed during an X-class flare in EUV and soft X-ray bands, with the EUV QPPs leading the soft X-ray QPPs by approximately a quarter of a period.
Kupriyanova et al. \cite{2019MNRAS.483.5499K} found that ~1-min QPPs in the temperature and emission measure perturbations during the decay phase of an X-class flare appear to be approximately in anti-phase with each other. This behaviour was interpreted as evidence for the second spatial harmonic of a standing slow magnetoacoustic mode controlling a quasi-periodic energy relaxation process in the post-flare arcade. Phase shifts between ~40-s QPPs in EUV, hard and soft X-rays, and radio bands have been reported in \cite{2023Univ....9..215X}. 

In this paper, we present a multi-wavelength analysis of QPPs observed during an X-class solar flare, in which a common $\sim$5-minute periodicity is detected across EUV and X-ray bands. We isolate the QPP signatures in individual EUV channels and identify systematic phase drift between the RHESSI HXR and EUV emissions.

The structure of the paper is as follows. In Section~\ref{Sec:Obs} we describe the observational data and their pre-processing. Section~\ref{Sec:Res} presents the analysis and results of the detected QPPs. Section~\ref{Sec:Con} discusses the implications of our findings and summarizes the main conclusions.

\section{Observations}
\label{Sec:Obs}

\begin{figure}[h!]
\centering
\includegraphics[width=\textwidth]{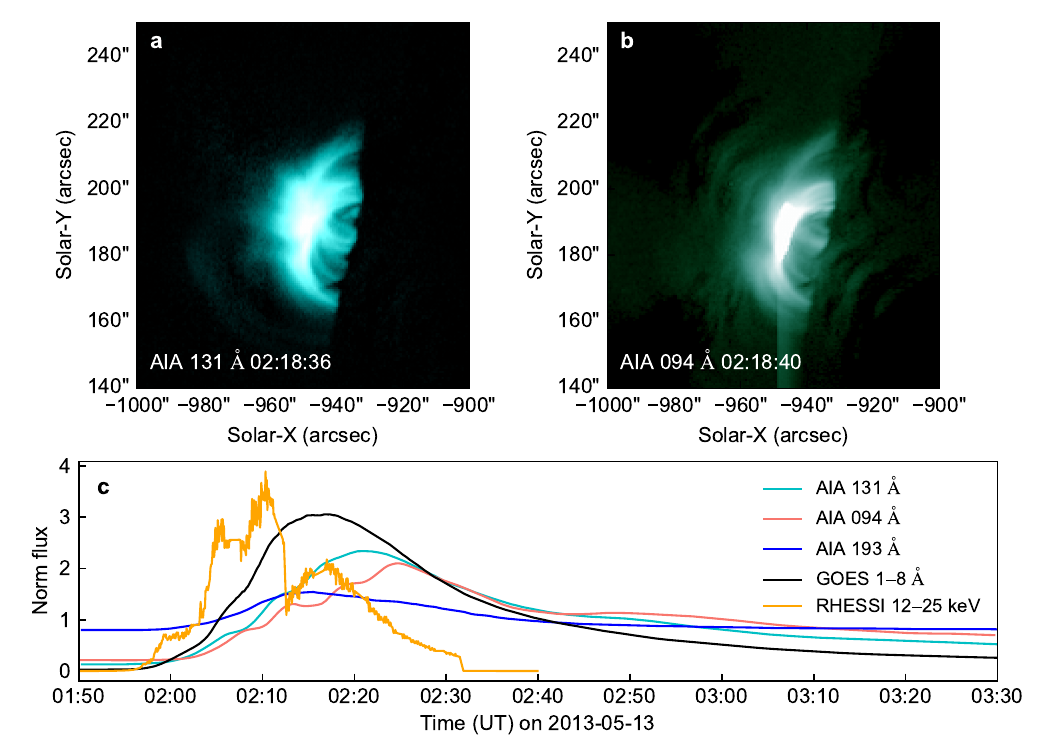}
\caption{{Multi-wavelength observations of QPPs during the X1.7 flare on 2013 May 13 in NOAA AR 11748.} (a) and (b) AIA 131~\AA{} and 094~\AA{} images of the flare. (c)
Raw integrated light curves from SDO/AIA (131~\AA{}, 094~\AA{}, and 193~\AA{}), GOES (1--8~\AA{}), and RHESSI (12--25 keV).
\label{fig:light_curve}
}
\end{figure}

On 2013 May 13, a GOES X1.7-class solar flare occurred in NOAA active region (AR) 11748, which was located close to the eastern limb of the Sun. The limbward position of the active region provided a favorable side-on viewing geometry, allowing the large-scale coronal structure associated with the flare to be clearly identified. The flare exhibited a rapid impulsive phase followed by a long-lasting gradual phase, with elevated emission persisting for more than one hour (Figure~\ref{fig:light_curve}).
NOAA AR 11748 was among the most flare-productive active regions of Solar Cycle 24 and produced several X-class flares during 2013 May 13–15. Previous studies have extensively investigated QPPs associated with the X3.2 flare on 2013 May 14 originating from this active region, using X-ray, microwave, and EUV observations (e.g., \cite{2015A&A...574A..53K,2015AdSpR..56.2769C,2017ApJ...836...84D}). In contrast, our study focuses on the earlier X1.7 flare on 2013 May 13, which has received comparatively less attention in the context of QPP studies.

We used observations from the Atmospheric Imaging Assembly (AIA) onboard the Solar Dynamics Observatory (SDO) \cite{2012SoPh..275...17L,2012SoPh..275....3P}. AIA provides full-disk solar images in multiple EUV passbands with a spatial resolution of $\sim1.5''$ and a nominal cadence of 12~s, enabling detailed investigations of coronal plasma evolution over a wide temperature range. In this work, we primarily analyze the AIA 131~\AA{}, 94~\AA{} and 193~\AA{} channels,  During large flares, these passbands become highly sensitive to hot flare plasma, the characteristic temperatures for these passbands, along with the 304~\AA{} and 335~\AA{} channels examined for comparison, are summarized in Table~\ref{tab:periods}. The AIA data were processed using the aia\_prep routine in the SolarSoft (SSW)/IDL module, following standard calibration procedures \cite{1998SoPh..182..497F,2012SoPh..275...17L}. These steps include the removal of the CCD bias and dark current, correction for flat-field effects, normalization by exposure time, and conversion to physical units. 

During the impulsive and early gradual phases of the flare,  the AIA Automatic Exposure Control (AEC) mechanism was triggered, leading to variable exposure times. To ensure consistent temporal sampling, we exclusively selected frames with a uniform nominal exposure time, resulting in an effective cadence of 24~s. Light curves were constructed by integrating the EUV intensity over a region of interest (ROI) centered at ($-1000^{\prime\prime}$, $200^{\prime\prime}$) in helioprojective coordinates,  with a size of $400^{\prime\prime} \times 400^{\prime\prime}$, which encompasses the saturated core of the flare. This approach preserves the relative temporal modulation of the signal, which is the primary focus of the present QPP analysis.
  
Soft X-ray fluxes were obtained from the GOES 15/XRS instrument in the 0.5--4~\AA{} and 1--8~\AA{} channels. Both GOES soft X-ray channels exhibit highly consistent temporal evolution throughout the flare.  A cross-correlation analysis reveals that their normalized light curves are almost identical (with a correlation coefficient of $\sim$0.97), indicating that the choice between the two channels does not affect the QPP analysis. Therefore, we representatively focus on the 1--8~\AA{} channel. The GOES data offer continuous full-Sun coverage with a native time resolution of 2~s, making them well suited for investigating flare-associated temporal variability. For consistency with the RHESSI hard X-ray observations, the GOES 1--8~\AA{} flux was re-binned to a time resolution of 4~s. Hard X-ray observations were obtained from the Reuven Ramaty High Energy Solar Spectroscopic Imager (RHESSI) \cite{2002SoPh..210....3L}. We analyze the RHESSI 12--25 keV energy band,  which captures the highest-energy processes during the impulsive phase, comprising contributions from both super-hot thermal bremsstrahlung and non-thermal electron precipitation \cite{2017ApJ...845L...1G}. The RHESSI light curves used in this study were generated following standard data reduction procedures, including detector selection, decimation correction, and background subtraction, as described in previous flare QPP studies (e.g., \cite{2016ApJ...833..284I,2017ApJ...836...84D}). The resulting time series provide complementary constraints on the high-energy emission associated with the flare.

\section{Results}
\label{Sec:Res}

\subsection{Dominating periodicities}
\label{Sec:Res_per}

We first investigate the characteristic periodicity of the QPP during the impulsive phase (01:50--02:40~UT) of the X-class flare. To avoid introducing artificial high-frequency components through detrending procedures, and to preserve the intrinsic temporal properties of the signal, we begin our analysis using the raw flux time series obtained from the AIA 131~\AA{} channel and the RHESSI 12--25~keV energy band. For each time series, we compute the power spectral density (PSD) using the fast Fourier transform (FFT). The statistical significance of spectral peaks is then assessed against a red-noise background model by means of a Markov Chain Monte Carlo (MCMC) approach.

\begin{figure}[h]
\centering
\includegraphics[width=\textwidth]{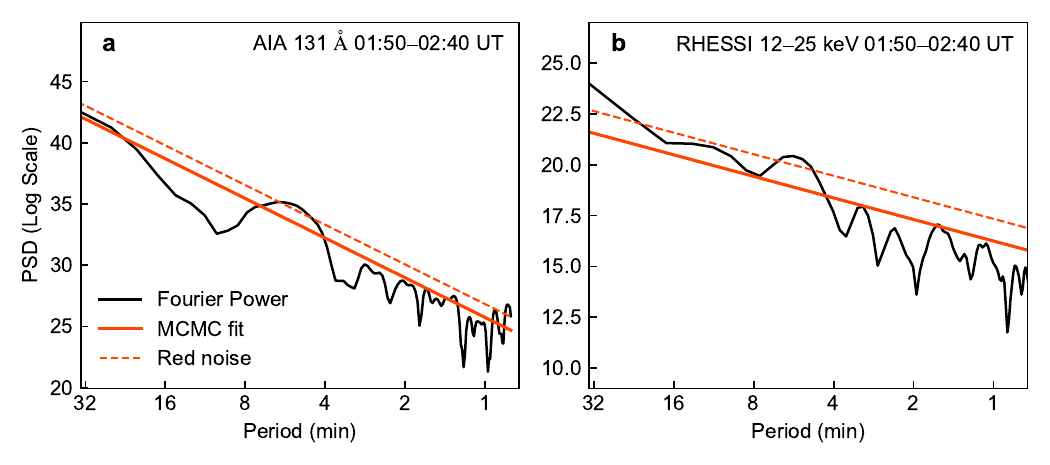}
\caption{Fourier power spectral analysis of QPPs during the impulsive phase of the flare. The power spectra were derived from the undetrended light curves shown in Figure~\ref{fig:light_curve}a. Panel (a) displays the power spectrum of the AIA 131~\AA{} signal, and (b) shows the RHESSI 12--25~keV signal spectrum. The red line represents the background red-noise model fitted using a Markov Chain Monte Carlo (MCMC) algorithm, and the dashed red line marks the 95\% confidence level.
\label{fig:mcmc}
}
\end{figure}

The resulting PSDs and their corresponding MCMC fits are shown in Figure~\ref{fig:mcmc}. The dominant QPP period is estimated from the maximum of the background-normalized power spectrum within the predefined period range. The associated uncertainty is estimated empirically from the half-width at half-maximum of the normalized spectral peak. In both the AIA~131~\AA{} and RHESSI 12--25~keV time series, the dominant periodic component is detected above the 95\% confidence level, with periods of $5.12 \pm 1.04$~min and $5.69 \pm 0.79$~min, respectively. A comparable periodicity is also identified in the AIA~94~\AA{} light curve, where a spectral peak at $5.25 \pm 1.01$~min exceeds the significance threshold. Fourier-based techniques are known to be poorly applicable to non-stationary signals typical for QPPs. However, as the observed oscillatory patterns do not show a visible modulation of the oscillation period, the signal non-stationarity leads to the broadening of the Fourier spectral peaks, while keeping the approximately correct location of the maximum. As the spectral peaks obtained for the amplitude-modulated signal exceed the detection threshold, we accept the outcome of the analysis.

In contrast, the raw GOES 1--8~\AA{} and AIA 193~\AA{}, 335~\AA{}, and 304~\AA{} power spectra are dominated by low-frequency red noise from the slowly varying background emission, and no peak exceeds the 95\% significance level within the period range of interest. For consistency, all channels are analyzed using the raw light curves only, without applying any additional detrending that might introduce otherwise undetected periodic signals. Accordingly, no dominant period is reported for these channels in Table~\ref{tab:periods}.

The close agreement between the EUV and hard X-ray periodicities indicates that the QPP manifests coherently in both thermal and high-energy emission during the flare. Moreover, the independent and simultaneous detection of the same periodic signal with two different instruments rules out an instrumental or artificial origin of the observed periodicity.

\begin{table}[!h]
\centering
\caption{Summary of dominant periods detected in different wavebands from the raw light curves. Period uncertainties are empirical estimates based on the half-width at half-maximum of the background-normalized spectral peak. The listed temperatures are representative values adopted for the dominant flare-related response of each channel \cite{2010A&A...521A..21O}.}
\label{tab:periods}
\small
\begin{tabular}{lcccc}
\hline
Instrument / Channel & Data      & Dominant         & Char.     & Note \\
                     & Treatment & Period (min)     & Tem. (MK) & \\
\hline
AIA 131~\AA{}        & Raw & $5.12 \pm 1.04$     & 11.2 & Significant\\
AIA 94~\AA{}         & Raw & $5.69 \pm 0.79$     & 7.1 & Significant\\
RHESSI 12--25~keV    & Raw & $5.25 \pm 1.01$     & N/A & Significant\\
GOES 1--8~\AA{}      & Raw & --                  & Multi-thermal& No significant peak \\
AIA 193~\AA{}        & Raw & --                  & 17.8 & No significant peak\\
AIA 335~\AA{}        & Raw & --                  & 2.8 & No significant peak\\
AIA 304~\AA{}        & Raw & --                  & 0.05& No significant peak \\
\hline
\end{tabular}
\vspace*{-4pt}
\end{table}

Having established the presence of a statistically significant periodic signal, we next isolate the oscillatory component in order to examine its temporal evolution. To minimize contamination from long-term trends while avoiding the introduction of spurious oscillations, we remove the low-frequency background using a smoothing window substantially longer than the detected QPP period. Specifically, a window length of 700~s is adopted, and the background trend is estimated using a third-order Savitzky--Golay filter \cite{1964AnaCh..36.1627S}.
Sensitivity tests with nearby window lengths indicate that the detrended oscillatory pattern is not qualitatively affected. The detrended light curves are then obtained by subtracting the smoothed background from the original time series, following procedures commonly employed in previous QPP studies (e.g. \cite{2013PASJ...65S...3K, 2019ApJ...886L..25Y}). 

\begin{figure}[h]
\centering
\includegraphics[width=0.5\textwidth]{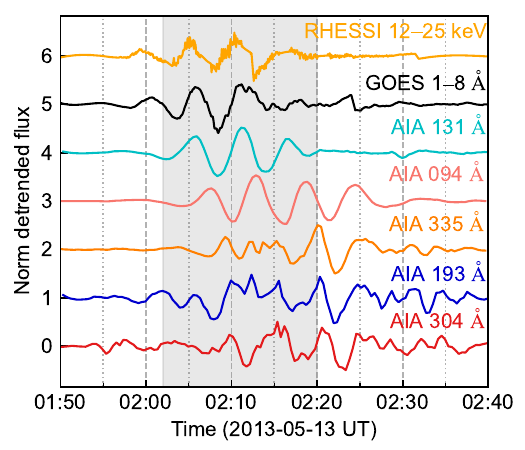}
\caption{Detrended and normalized light curves from the AIA, GOES, and RHESSI channels analyzed in this study, covering 01:50--02:40~UT on 2013 May 13. The background trend was estimated with a 700~s smoothing window and removed from each light curve. The shaded region indicates the interval 02:02--02:20~UT used for the subsequent detailed analysis. No additional band-pass filtering is applied in this figure. The curves are vertically offset for clarity.
\label{fig:detrend_curve}
}
\end{figure}

Figure~\ref{fig:detrend_curve} presents the detrended light curves of seven wavebands, including five AIA EUV channels (131~\AA{}, 94~\AA{}, 193~\AA{}, 335~\AA{}, and 304~\AA{}), the GOES 1--8~\AA{} soft X-ray flux, and the RHESSI 12--25~keV hard X-ray emission. After removal of the low-frequency background, quasi-periodic pulsations remain clearly visible in the main wavebands, demonstrating that the detected oscillatory signals are not artifacts introduced by long-term trends. Following detrending,  the main QPP-bearing light curves exhibit an oscillatory component with a characteristic timescale of approximately 5 minutes. The QPPs persist for about three to five cycles during the impulsive and early decay phases of the flare. The oscillations are non-stationary in amplitude, showing a pronounced modulation envelope. No clear monotonic drift of the dominant period is observed.

The oscillatory signals seen in the 131~\AA{}, 94~\AA{}, 1--8~\AA{}, and 12--25~keV wavebands appear to be similar, although phase shifted. The AIA 193~\AA{} signal is more complex.  In the detrended AIA 193~\AA{} light curve, an additional component near 3~minutes is present besides the $\sim$5-min QPP, and the corresponding detrended power spectrum shows two significant peaks. To facilitate the subsequent comparison of the 5-min signal with those in the other channels, the AIA 193~\AA{} light curve is further band-pass filtered in the 4--8~minute range in the later analysis, thereby suppressing the 3-min component. The curves shown in Figure~\ref{fig:detrend_curve} are detrended only and are not band-pass filtered.

\subsection{Phase shifts}
\label{Sec:Res_phase}

\begin{figure}[h]
\centering
\includegraphics[width=1\textwidth]{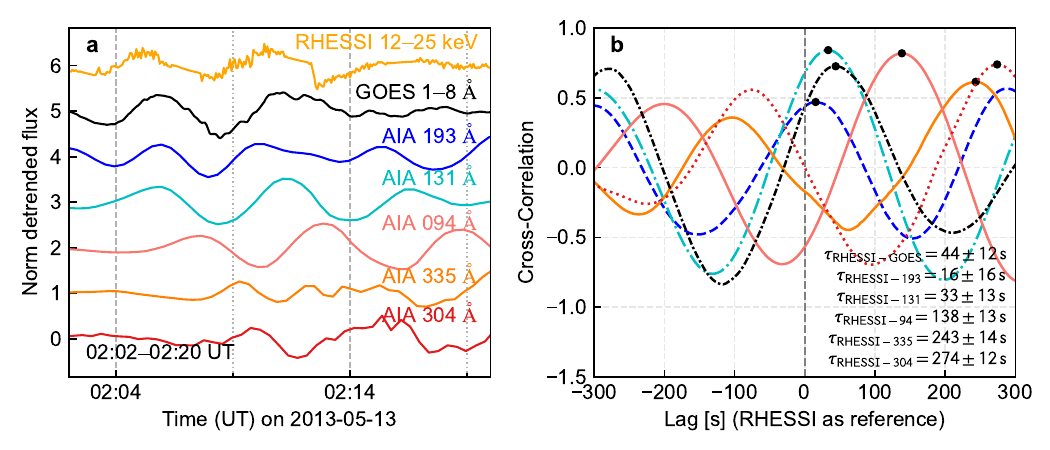}
\caption{
Cross-correlation analysis of the detrended QPP signals in the AIA 193~\AA{}, 131~\AA{}, 94~\AA{}, 335~\AA{}, and 304~\AA{} channels, GOES 1--8~\AA{}, and RHESSI 12--25~keV. 
(a) Enlarged view of the detrended and normalized light curves over the interval 02:02--02:20~UT, highlighted in Figure~\ref{fig:detrend_curve}. 
(b) Cross-correlation functions between the RHESSI 12--25~keV signal and the other wavebands, computed over the interval shown in panel~(a). The black dots mark the lag corresponding to the maximum correlation in each case.
\label{fig:cross_corrlation}
}
\end{figure}

To quantify the relative timing between the 5-min oscillatory signals observed in different wavebands, we perform a cross-correlation analysis between pairs of detrended time series. The cross-correlation provides a direct measure of the temporal offset between correlated oscillatory signals. In particular, this approach has been widely used to investigate the temporal evolution and cooling behavior of coronal plasma (e.g. \cite{2012ApJ...753...35V,2019ApJ...880...56B}). For the cross-correlation analysis, all light curves were interpolated onto a common 4 s time grid, this interpolation was used only for uniform lag estimation.
Figure~\ref{fig:cross_corrlation} examines the temporal relationships among the detrended and normalized light curves shown in Figure~\ref{fig:detrend_curve}, focusing on the time interval displayed in panel~(a). Using the 12--25~keV signal as a reference, we compute the cross-correlation functions (CCFs), i.e., the dependence of the cross-correlation coefficients between this signal and the signals detected in other channels on the time lags between the signals. The uncertainties of these time lags are estimated via a Monte Carlo flux randomization approach over iteration.

Panels~(a) and (b) present the detailed analysis of the interval 02:02--02:20~UT highlighted in Figure~\ref{fig:detrend_curve}. As shown in panel~(b), the RHESSI-referenced CCFs exhibit well-defined maxima for all channels, indicating a coherent quasi-periodic modulation across the different emissions. Taking the RHESSI 12--25~keV hard X-ray emission as the temporal reference, the AIA 193~\AA{} channel is nearly synchronous with RHESSI, with a formally estimated lag of $\tau_{\mathrm{RHESSI-193}} = 16 \pm 16~\mathrm{s}$. Because this lag is shorter than the native AIA cadence of 24~s, it should be interpreted cautiously rather than as a precisely delay.
The AIA 131~\AA{} channel and the GOES 1--8~\AA{} soft X-ray flux show modest delays of 
$\tau_{\mathrm{RHESSI-131}} = 33 \pm 13~\mathrm{s}$ and 
$\tau_{\mathrm{RHESSI-GOES}} = 44 \pm 12~\mathrm{s}$, respectively. Larger delays are found in the AIA 94~\AA{}, 335~\AA{}, and 304~\AA{} channels, with $\tau_{\mathrm{RHESSI-94}} = 138 \pm 13~\mathrm{s}$, $\tau_{\mathrm{RHESSI-335}} = 243 \pm 14~\mathrm{s}$, and $\tau_{\mathrm{RHESSI-304}} = 274 \pm 12~\mathrm{s}$, respectively.

Specifically, expressed in phase units assuming a 5.5-min period, the QPPs in the AIA 193~\AA{}, 131~\AA{}, and GOES 1--8~\AA{} emissions lag the QPP in the RHESSI 12--25~keV channel by $\sim 0.05$, $\sim 0.10$, and $\sim 0.13$ cycles, respectively. Larger phase shifts are found for the AIA 94~\AA{}, 335~\AA{}, and 304~\AA{} channels, which are delayed by approximately $\sim 0.42$, $\sim 0.74$, and $\sim 0.83$ cycles relative to the hard X-ray signal, for AIA 335~\AA{} and 304~\AA{}, these values are close to one oscillation cycle and are therefore subject to phase-wrapping ambiguity.

Hence, these results reveal a clear and systematic phase ordering during the $\sim$5.5~min QPP interval: the RHESSI 12--25~keV emission responds earliest, followed by the AIA 193~\AA{} and 131~\AA{} channels and then the GOES 1--8~\AA{} soft X-rays, while substantially larger delays are found in the AIA 94~\AA{}, 335~\AA{}, and 304~\AA{} channels. This ordered phase progression is consistent with a scenario in which the quasi-periodic energy release is manifests first in the non-thermal and hotter flare-sensitive emissions, whereas the responses in the other channels appear later. Although the AIA 335~\AA{} and 304~\AA{} channels show similarly large delays, the 304~\AA{} emission should be interpreted with caution in a temperature-dependent context, as it is not a simple diagnostic of hot coronal flare plasma and may include substantial lower-atmosphere and transition-region contributions. For this reason, we do not include the 304~\AA{} channel in the main temperature-dependent fit.

\begin{figure}[h]
\centering
\includegraphics[width=0.5\textwidth]{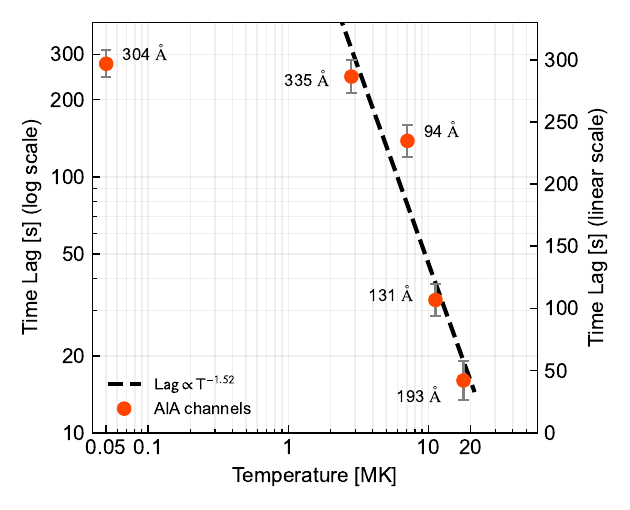}
\caption{
Time lag of the QPP signal in each waveband relative to the RHESSI 12--25~keV emission as a function of representative channel temperature. The temperatures shown on the x-axis correspond to the flare-related dominant response of each AIA passband \cite{2012SoPh..275...17L}. The left vertical axis is logarithmic, while the right vertical axis gives the corresponding linear scale for the error bars. The dashed line shows the best-fit power-law relation for the AIA 193~\AA{}, 131~\AA{}, 94~\AA{}, and 335~\AA{} channels. The AIA 304~\AA{} point is shown for reference only and is not included in the fit.
\label{fig:Tem}
}
\end{figure}

We also examined the relationship between the measured time lags, $\tau$, and the representative peak temperatures, $T_{\mathrm{peak}}$, of the AIA passbands, as shown in Figure~\ref{fig:Tem}. The temperatures used on the x-axis are representative values adopted for the flare-related dominant response of each AIA passband. Given the strong heating during this X-class flare, we adopt the high-temperature flare components for the 193~\AA{}, 131~\AA{}, and 94~\AA{} channels, following the instrumental temperature response summarized by \cite{2012SoPh..275...17L,2010A&A...521A..21O}. The specific representative peak temperatures ($T_{\mathrm{peak}}$) adopted for each channel in this analysis are summarized in Table~\ref{tab:periods}.

The characteristic conductive cooling timescale of a coronal loop scales as $\tau_{\mathrm{cond}} \propto nL^{2}T^{-5/2}$ in analytical loop-cooling models \cite{1994ApJ...422..381C,2005psci.book.....A}. If the observed inter-band QPP lags are interpreted as differences in cooling time, the lag is likewise expected to scale as $\tau \propto T^{-5/2}$. We therefore fitted the measured lags in log--log space using the AIA 131~\AA{}, 193~\AA{}, 94~\AA{}, and 335~\AA{} channels (Figure~\ref{fig:Tem}), and estimated the uncertainty of the fitted power-law index with Monte Carlo realizations by perturbing the measured lags within their $1\sigma$ errors in log space while keeping the adopted channel temperatures fixed.  The fit yields $\alpha = -1.52 \pm 0.31$ with $R^2 = 0.91$, indicating an inverse lag--temperature relation in which cooler channels respond later. However, the inferred slope is shallower than the simple conductive-cooling prediction \cite{2005psci.book.....A}, suggesting that conductive cooling alone may not fully explain the observed trend.

\section{Conclusions}
\label{Sec:Con}

In this study, we report the detection of a $\sim$5-min QPP during an X-class solar flare, observed in EUV and X-ray wavebands. The periodic variations in the EUV and X-ray emission intensities exhibit a nearly harmonic form. The simultaneous detection of the QPP in different bands with different instruments excludes the artificial nature of the periodicity. The QPP manifests as a short-lived, non-stationary wave packet, persisting for only a few cycles. A central result of this work is the presence of systematic phase drifting between the oscillatory patterns detected in different observational channels.

This temperature-dependent phase ordering suggests that the leading RHESSI signal is associated with the highest-temperature component in the system. We note, however, that neither the AIA EUV channels nor the RHESSI 12--25~keV band is a purely single-temperature diagnostic. The AIA 94, 131, and 193~\AA{} passbands can contain contributions from both hot and cool plasma components, while the RHESSI 12--25~keV emission may also include a significant thermal contribution in X-class flares \cite{2010A&A...521A..21O,2011A&A...535A..46D,2017ApJ...845L...1G}. Nevertheless, our interpretation is based primarily on the relative timing and systematic phase ordering among the channels. In this context, such mixed temperature responses do not alter the main result that the QPP signals exhibit a coherent temperature-dependent delay pattern.

The finding that QPPs associated with plasmas at different temperatures exhibit nearly identical oscillation periods argues against an interpretation of the oscillations in terms of standing slow magnetoacoustic waves. For a given coronal loop length, the periods of standing slow magnetoacoustic waves increase with temperature. Perhaps the 5-min periodicity is imposed by repetitive magnetic reconnection, periodically triggered by the leakage of 5-min oscillations from the lower layers of the solar atmosphere into the corona along magnetic field lines. This scenario is similar to the modulation of flaring emission by chromospheric 3-min oscillations observed in \cite{2009A&A...505..791S}. This interpretation is consistent with the presence of QPPs in the 12--25~keV emission, which are associated with non-thermal electrons. The source of the 5-min periodicity may be chromospheric 5-min oscillations localised at the umbra-penumbra boundary of the sunspot \cite{2014ApJ...792...41Y}. The association of QPPs with magnetic reconnection has been indicated by the simultaneous detection of QPPs and recurrent jets sharing the same periodicity \cite{2022FrASS...932099L, 2023ApJ...945..113M}.

The phase shifts between QPPs detected in thermal emissions, i.e., in the 193~\AA{}, 131~\AA{}, 94~\AA{}, and 1--8~\AA{} channels, which are associated with plasmas at different temperatures, may be attributed to temperature variations in the emitting plasma. The finding that the non-thermal 12--25~keV QPP signal precedes the thermal QPPs suggests that the plasma is periodically heated by magnetic reconnection. The subsequent temporal evolution of the thermal QPP signatures therefore provides important constraints on the dominant cooling and energy transport processes following each heating episode. 

This indicates that, after the cessation of non-thermal electron beam injection,  the subsequent plasma evolution may be largely governed by conductive cooling. This interpretation is broadly consistent with the radiative hydrodynamic simulations of Kerr et al. \cite{2026NatAs..10..202K}, who showed that the post-heating atmospheric response depends sensitively on the subsequent thermodynamic evolution of the atmosphere, including conductive cooling after the cessation of particle-beam heating. In this framework, our observations can be understood as revealing, in the time domain, a cyclic process of impulsive non-thermal energy injection followed by conduction-dominated cooling: periodic magnetic reconnection modulates the acceleration and precipitation of non-thermal electrons, giving rise to the HXR QPPs, while the subsequently heated plasma cools under the dominance of thermal conduction and becomes visible sequentially in cooler channels, producing the observed temperature-dependent phase shifts of the thermal QPPs.

In this study, we do not account for the time variation of the plasma density and associated flows. Furthermore, it is unclear why the detected QPP patterns exhibit a nearly harmonic shape, since a relaxation or a sawtooth-like temporal profile might be expected instead. In addition, the 1--8~\AA{} QPP exhibits a larger time delay relative to the nonthermal signal than those observed in the cooler 193~\AA{} and 131~\AA{} channels. This discrepancy may be caused by the influence of observational noise, which affects the accuracy of the phase-shift determination. This interpretation would benefit from measuring the density and temperature of the emitting plasma which, in principle, could be done by the differential emission measure estimated using multiple observational wavelengths. However, in this event such an estimation turned out to be impossible because of the saturation of the flare core in the SDO/AIA observations. Validation of the suggested scenario requires dedicated forward modelling of the evolution of observables, following the procedure introduced, in particular, in \cite{2016FrASS...3....4V, 2021A&A...656A.120F} with image desaturation \cite{2019ApJ...882..109G,2021SoPh..296...56Y}. These results are based on a single X-class flare event, and further studies of a larger sample are required to assess their general applicability.

\ack{LBF, DY, SS and ET are supported by National Natural Science Foundation of China (NSFC, 11803005), the Guangdong Natural Science Funds for Distinguished Young Scholar (2023B1515020049), the Shenzhen Science and Technology Project (JCYJ20240813104805008), and the Science and Technology Program of Guangdong Province (grant No. 2025B1212050001). VMN is supported by ERC grant 101201424 (ACDCSUN). Views and opinions expressed are however those of the author(s) only and do not necessarily reflect those of the European Union or the European Research Council Executive Agency. Neither the European Union nor the granting authority can be held responsible for them.}

\bibliographystyle{RS} 
\bibliography{fu_qpp_2025}%

\end{document}